# The Arc of the Data Scientific Universe

A comment on Sabina Leonelli's "Data Science in Times of Pand(dem)ic"

David Leslie, The Alan Turing Institute
dleslie@turing.ac.uk

## Abstract

In this paper explore the scaffolding of normative assumptions that supports Sabina Leonelli's implicit appeal to the values of epistemic integrity and the global public good that conjointly animate the ethos of responsible and sustainable data work in the context of COVID-19. Drawing primarily on the writings of sociologist Robert K. Merton, the thinkers of the Vienna Circle, and Charles Sanders Peirce, I make some of these assumptions explicit by telling a longer story about the evolution of social thinking about the normative structure of science from Merton's articulation of his well-known norms—those of universalism, communism, organized skepticism, and disinterestedness—to the present. I show that while Merton's norms and his intertwinement of these with the underlying mechanisms of democratic order provide us with an especially good starting point to explore and clarify the commitments and values of science, Leonelli's broader, more context-responsive, and more holistic vision of the epistemic integrity of data scientific understanding, and her discernment of the global and biospheric scope of its moral-practical reach, move beyond Merton's schema in ways that effectively draw upon important critiques. Stepping past Merton, I argue that a combination of situated universalism, methodological pluralism, strong objectivity, and unbounded communalism must guide the responsible and sustainable data work of the future.

## 1. Introduction

The arc of the data scientific universe is long, but it bends toward the global public good. So Sabina Leonelli would seem to exhort in her masterful contribution ("Data Science in Times of Pand(dem)ic," this issue) to reconceptualizing the practice of data science in a post-COVID world. For Leonelli, good data science in the times of pandemic, and beyond, is not merely robust, reliable, and sound data science (epistemically speaking); it is publicly oriented data science (normatively speaking)—data science for the social good.

Leonelli's mapping of a new world of data scientific imaginaries where "collaborative, engaged forms of data work…seek to understand social and environmental needs, evaluate research directions and construct appropriate tools in dialogue with relevant communities" weaves together these epistemic and normative threads into a kind of forward-thinking cartographic tapestry. She stresses the need for "dedicated global infrastructures" that enable multilateral, interdisciplinary, and transnational networks of collaboration. This would help data-intensive research to widen its epistemic lens and integrate quantitative measurements and inferences with qualitative, domain-embedded, and context-sensitive observations and insights. Such a lengthened purview would expand its capacity to leverage causal understandings of complex and rapidly evolving pandemic environments and to produce more coherent and holistically informed solutions. The broader field of vision opened up by an interdisciplinary and global data scientific apparatus would also entail a widening (and sharpening) of its normative lens. By extending its moral-practical sightlines, this sort of inclusive data work could move beyond the myopic embrace of technological solutionism, algorithmic silver bullets, and gadget-based technocratic quick fixes and instead engage in the kind of long-term, complex analysis and problem-solving that remains responsive to changing social and environmental challenges.

Such a value-centred and whole-of-society approach to data science would be able to zoom in to address immediate pandemic-related problems such as the disparate effects of COVID-19 on minority ethnicities, vulnerabilities generated by environmental racism, and downstream economic impacts like energy poverty. It would likewise demand that data scientists and researchers remain socially attuned, community-involved, and self-critical—and that they engage in vigilant reflection about the potential impacts of their work on affected individuals and wider society. At the same time, this approach to data work would be able to zoom out and assume global and transnational perspectives that transcend predominant statist geopolitics and that develop a more biospheric vision focused on the micro- and macro-scale determinants of planetary health and sustainability. In a memorable passage, Leonelli makes this appeal for a recalibration and expansion of the normative vision of data work explicit: "*Data science needs to stop* *feigning neutrality, and instead work collaboratively with domain experts and relevant communities toward forging socially beneficial solutions*" (emphasis in original).

The significant energies that Leonelli dedicates to this kind of prospective mapmaking are largely intended to inspire and guide a rising generation of data scientists to see themselves as more than just mathematical guns-for-hire—more than mere mercenaries of a "'tool discipline' ready to serve any master." Data science—or *any science*, for that matter—starts from a particular *ethos*, a particular background of action-orienting norms, standards, commitments, and mores. Taken together, these form a kind of normative *habitus* (or structuring of propensities and tendencies) that sets in place and gives shape to the basic values, dispositions, and character traits that determine what good looks like in any given community of practice. Leonelli's implicit claim is that the ethos of responsible and sustainable data science and the habitus of its practices involve a particular set of commitments and values that are dually oriented to epistemic integrity (i.e., to providing as full and inclusive yet reflexive and critical a scientific understanding of the problem at hand as possible) and to the good of humanity and planetary life as a whole. This is why, she claims, "data scientists need to abandon the myth of neutrality that is attached to a purely technocratic understanding of what data science is as a field—a view that depicts data science as the blind churning of numbers and code, devoid of commitments or values except for the aspiration toward increasingly automated reasoning." Where data scientists, who view themselves simply as socially disembodied, quantitative analysts, engineers, or code-churners go wrong is that they are insufficiently attentive to the commitments and values that undergird the integrity of their knowledge practices and the ethical permissibility of the projects, enterprises, and use-contexts in which they involve themselves.

In this comment, I want to explore the scaffolding of normative assumptions that supports Leonelli's implicit appeal to the values of epistemic integrity and the global public good that conjointly animate the ethos of responsible and sustainable data work. Drawing primarily on the writings of sociologist Robert K. Merton, the thinkers of the Vienna Circle, and Charles Sanders Peirce, I hope to make some of these assumptions explicit by telling a longer story about the evolution of social thinking about the normative structure of science from Merton's articulation of his well-known norms—those of universalism, communism, organized skepticism, and disinterestedness—to the present. I will show that while Merton's norms and his intertwinement of these with the underlying mechanisms of democratic order provide us with an especially good starting point to explore and clarify the commitments and values of science, Leonelli's broader, more context-responsive, and more holistic vision of the epistemic integrity of data scientific understanding, and her discernment of the global and biospheric scope of its moral-practical reach, move beyond Merton's schema in ways that effectively draw upon important critiques.

## 2. Crisis Invites Self-Appraisal

In 1942, amid transatlantic reverberations of the waves of anti-intellectualism and racist pseudoscience that had come to dominate the political culture of Nazi Germany, a young American sociologist named Robert Merton set out to sketch a picture of what he termed "the ethos of science," namely, "that affectively toned complex of values and norms which is held to be binding" on any scientist worthy of the title (Merton, 1942/1968, p. 605).[1]

Merton's intervention, published originally as an article entitled "Science and Technology in a Democratic Order," opened with a motivation. Whereas, for the many generations nourished by the fruits of the modern scientific revolution, faith in the culture of science was "unbounded, unquestioned, unrivalled" (Merton, 1942/1968, p. 604), the explosion of revolts from science in the mid-20th century was enough to give even the most devoted scientist pause. "Incipient and actual attacks upon the integrity of science," he wrote, "have led *scientists to recognize their dependence on particular types of social structure*" (Merton, 1942/1968, p. 604, emphasis in original). As a kind of social practice—a form of social activity—science, like any other social practice involving norms, interactions, obligations, and interests, was conditioned by the institutional configurations and constraints of the existing societal order simultaneously as it was subject to the winds of cultural and political change. For Merton, this meant a need for those committed to the institution of science to engage in critical self-evaluation: "An institution under attack must re-examine its foundations, restate its objectives, seek out its rationale. Crisis invites self-appraisal" (Merton, 1942/1968, p. 604).

---

[1] Merton had, in fact, directly engaged the problems of anti-intellectualism and racist pseudoscience in Nazi culture in an earlier 1938 essay titled "Science and Social Order," which can be seen as a stage-setting precursor to his 1942 examination of the normative structure of science (Merton, 1938/1968).

For his part, Merton pressed forward with this task of self-appraisal by carrying out a kind of reconstructive two-step. First, he explicated what he took to be the ethos of science—that internalized set of norms and values that, for him, at once fashioned "scientific conscience" and functioned as the body of institutional imperatives through which the epistemic and moral-practical integrity of science was instantiated. He listed these imperatives as universalism, communism, organized skepticism, and disinterestedness. Second, he unearthed the particular type of social structure, what he called "the democratic order," that enabled the ethos of science (and that was effectively integrated with it). Taking the word "democratic" here as a social rather than a specifically political concept,[2] Merton teased out the elective affinity between democratic social order and the enabling conditions of the integrity of modern science, emphasizing common aspects like inclusivity, dialogical openness, institutionalized criticism, and egalitarian reciprocity, opportunity, and participation.

---

[2] This wider conception of democracy as a social rather than political idea has its origins in Tocqueville but finds expression in the heritage of social theory that is downstream from him (Elster, 2009)

Taken in conjunction, Merton's imperatives and his intertwinement of these norms with the underlying mechanisms of democratic order provide us serviceable ingress into clarifying the normative assumptions that underpin Leonelli's appeal to the commitments and values of epistemic integrity and the good of humanity. Let me, then, first probe Merton's reconstruction of the "normative structure of science," putting it into in dialogue with his intellectual influences and subsequent critics and expanding upon its strengths and failings. This will allow me to conclude by capturing the implicit normative advancements Leonelli makes in her article.

## 3. Universalism Through the Technomodernist Looking-Glass

The first imperative Merton sees as underlying the ethos of science is universalism. On his account, universalism, first and foremost, has to do with scientific reasoning and explanation. This cognitivist focus on the validity conditions of scientific justification is perhaps unsurprising given that he defines the "institutional goal of science" as "the extension of certified knowledge" (Merton, 1942/1968, p. 606). What is 'universal' in this epistemic sense is any truth-claim that is "subjected to *preestablished impersonal criteria*: consonant with observation and with previously confirmed knowledge" (Merton, 1942/1968, p. 607, emphasis in original). The rational acceptability of scientific reasoning and explanation cannot depend, along these lines, on any personal or social attributes of the claimant. Rather, 'objectivity' is established by impersonal criteria of validity that preclude any sort of particularism that might impeach the generality of the explanation or result.

To make his assertion about the imperative of universalism stick, Merton has to spell out what exactly he means by "preestablished impersonal criteria of validity," for, devoid of this piece of the argument, the linchpin of "certified knowledge" would remain underspecified and unable to support the weight of scientific objectivity he places upon it. That is, the content of the norms needed to fulfill the institutional objectives of science would remain ambiguous and obscure. For this, Merton reels in a group of epistemic concepts that had emerged in the 1920s and 1930s among a group of scientists and philosophers in Vienna (now known simply as the Vienna Circle).[3] He defines knowledge as "empirically confirmed and logically consistent statements of regularities (which are, in effect, predictions)" (Merton, 1942/1968, p. 606). Accordingly, he cleaves the "methodological rationale" of the mores of scientific justification into empirical and logical strands: "The technical norm of empirical evidence, adequate, valid and reliable, is a prerequisite for sustained true prediction; the technical norm of logical consistency, a prerequisite for systematic and valid prediction" (Merton, 1942/1968, p. 606). For Vienna Circle thinkers like Otto Neurath, Moritz Schlick, and Rudolf Carnap,[4] this parsing of criteria of validity into empirical and logical aspects had formed the cornerstone of a wider program of unified science that has since become known as logical positivism or logical empiricism.

It is easy to see why the Vienna thinkers may have appealed to Merton. As children of the heroic period of Einstein, Planck, and Bohr and witnesses to the lingering atrocities of the First World War, the members of the Vienna Circle were roused by the leaps in progress being made in the experimental sciences in the wake of the rise of the new physics. At the same time, they were horrified at the rising tides of propagandistic mythology and racist pseudoscience that characterized the ethnocentric and nationalist ideologies mushrooming across interwar Europe. Against this background, they fashioned themselves as philosophical defenders of precision, objectivity, and universality in scientific reasoning and as destroyers of the swelling tendencies at all levels of Occidental culture toward mysticism, metaphysics, and fanaticism. The guiding impulse behind their self-ascribed charge to patrol the boundary between sense and nonsense was a desire to clarify what they called a "scientific world-conception"(Hahn et al., 1929/2012)—an approach to research and argumentation that championed the application of a method of reflective analysis to clarify and reinforce the valid ways of justification available in collaborative human efforts to share experience and to cope with the hardships of material existence.

---

[3] While Merton did not make direct reference to the Vienna Circle thinkers, his familiarity with this school of thought is evidenced in an article written a year before, "Karl Mannheim and the Sociology of knowledge" (Merton, 1941/1957).

[4] There are, of course, many important differences between the view of each of these thinkers, but, for our purposes here I will stress the identity of the Vienna Circle as a whole—making distinctions between the perspectives of Neurath, Carnap, and Schlick where appropriate.

Such an approach involved, for them, a kind of two-phase process of problem-driven discovery. First, in order to set "the limits for the content of legitimate science" (Hahn et al., 1929/2012), "logico-empirical analysis," as they called it, had to meticulously dissect the experimental method of reasoning into logical and observational modes of judgment. This utilization of what is known as Hume's fork aimed to eliminate pseudostatements and to obtain building-blocks of meaning for scientifically valid communication by identifying correct paths of inference deriving *either* from tautologically sound 'relations of ideas' *or* from empirically testable 'matters of fact.' Second, in order to facilitate a practically effective recombination of the constructive tools of logic and the confirmatory tools of empiricism, the method called for a "new kind of scientific synthesis" (Neurath, 1937/1987, p. 139). This required establishing an actionable continuum between the categorical framework of the formal (logico-mathematical) or theory language and the baseline certainties of the observational language. The standards of measurement and combination stemming from the axioms of the former could, as it were, simply pass through in cooperative experimental practices to the hypothesis testing and empirical confirmation requisite for the corroborative assurances provided by the latter.

Merton's adoption of the technical norms of empirical evidence and logical consistency as fulcrums of the methodological rationale of science tapped into an informative ambiguity in the Vienna Circle's scientific world conception—one that put certain critical and progressive dimensions of its perspective in tension with some of its own tendencies toward scientism and epistemic reductivism. From the critical and progressive side, a notion of universalism that was anchored in criteria validity and processes of justification rather than in the content of truth-claims (that corresponded to an external reality in which that content could be found) signaled the adoption of a kind of epistemic humility and self-restraint (Leslie, 2020, Appendix). Because modern scientific reasoning was no longer entitled to gain knowledge about the world by helping itself to an inherent order of things, what came to count as rational was its ability to "solve problems successfully through procedurally suitable dealings with reality" (Habermas, 1992). Neurath, in this vein, espoused the value of procedural rationality and celebrated the disposal of "metaphysics with its senseless formulations." He urged that scientists disavow "the *material* mode of speech and use the formal mode (cf. Carnap), which does not speak of the *presence of certain entities* but of the *legitimacy of sentences*" (Neurath, 1932/1987, p. 8).

This deflationary impetus led logical empiricists like Neurath to conceive of the universal character of science *in the social and pragmatic sense* of a unifying practice of sharing experience and redeeming reasons in a common language—albeit one exclusively shunted into observational and analytical modalities of speech and expression. The universalism of the procedural mechanisms of validation and verifiability served as a collaborative response to science's demand for radical intersubjectivity even in the wake of the expanding pluralism of modern ways of thinking, acting, and being. It was a kind of concerted rejection of the exclusionary and divisive tendencies of intolerant metaphysical or religious outlooks, and it demanded—at social-structural and institutional levels—inclusive participation in the pursuit of scientific consensus and unmitigated, organized, and methodical self-criticism. As Neurath put it in his introductory essay to a series of Vienna Circle monographs published in the 1930s under the title *Unified Science*: "*Metaphysical terms divide—scientific terms unite*. Scientists united by a unified language form a kind of workers' republic of letters no matter how much else may divide them as men" (Neurath, 1932/1987, p. 23).

Seen in this light, the revolutionary project of the logical empiricism to burst the balloon of metaphysical self-inflation in culture, philosophy and science was driven by what Rudolf Carnap referred to in the preface to *The Logical Structure of the World* as a more fundamental and societally ramifying "life-stance" oriented to the pursuit of progressive social and cultural improvement through inexhaustible and tireless criticism as well as the slow, painstaking work of applied scientific research and innovation (Carnap, 1928/2003; see also, Awodey & Klein, 2004; Nelson, 2012). The crux of such a ceaseless and intergenerational undertaking was perhaps best captured in Neurath's famous image of the project of humanity as the burden to continuously and democratically rebuild the ship of civilization in the open seas of ontological uncertainty, no shores in sight. All of these normative dimensions of the ethos of modern science—from the importance of inclusive participation, egalitarian consensus-building, and universalistic solidarity to the institutionalization of organized self-criticism—Merton embraced.

There was, however, another, less tolerant and less progressive, side to Vienna Circle thinking that ended up at cross-purposes with these norms and that presaged limitations of Merton's own concept of universalism. Emboldened by the practical successes of the modern natural sciences and by a corresponding faith in the 'inexorable march of technology,' members of the Circle embraced what historian Peter Galison has called "technomodernism"—the "hard-edged, hard-boiled rationalism of technical objectivity" (Galison, 1993, p. 87). As has been emphasized, the Vienna Circle thinkers adopted a verificationist epistemology, that is, the belief that sources of knowledge and meaningfulness originated in either logical proofs, namely, statements that are shown to be true by definition, or empirically confirmable statements. Now, on its own, this understanding of the nature of scientific knowledge seems to commonsensically reflect the technical norms and criteria of validity that have funded the success of the modern exact sciences. But, the technomodernist impetus to scientism, namely, its privileging of natural scientific language, knowledge, and methods over those of other branches of learning and culture (Sorell, 2013), led the Vienna thinkers to prioritize and extend the scope of their verificationist epistemology across all domains of social practice. This meant a monistic and homogenizing reduction of the meaningfulness and significance of all the diverse coping vocabularies emergent within human experience (from those of the arts, ethics, and religion to those of history, the humanities, and the social sciences) to that of the exact, physical sciences. Along the lines of this scientistic *episteme*, it was the sovereign logic of scientific investigation and explanation that established what could be meaningfully said about the world and the 'unified science of physicalism' that alone was able to say anything meaningful at all. This veritably relegated the understandings and knowledge practices of the other branches of learning and culture to the dustbin of epistemic irrelevance.

Such an inflated scientistic self-understanding yielded other collateral weaknesses in logical empiricist thinking. Most crucially, the self-conceived ubiquity of quantitative and physicalist language caused a blind spot in verificationist epistemology that derived from its own evasion of the social, practical, and contextually anchored character of the activities of scientific investigation, observation, and justification. The desire for hard-edged technical objectivity meant that logical empiricist thinkers like Schlick and Carl Hempel all but eliminated the cognizing subject—the agent of scientific knowing—from their epistemic picture (de Regt et al., 2009). Scientific explanation, on this view, involved simply an *explanandum* (the phenomenon that was to be explained) and an *explanans* (the collection of theories or evidential premises and conclusions that was doing the explaining). The cognizing agent, whose values, commitments, and concerns animated and steered explanatory activity, was left out of the schema of scientific explanation altogether. So too were the contexts of the social practices and complex relations of research production that influenced and shaped the trajectories of scientific investigation.

Moreover, the observing scientific subject, who functioned at bottom, for logical empiricists, as the gatekeeper of the epistemic certainty of sensory experience, was treated merely as a disembodied vessel of reliable representation, an unproblematic "mirror of nature" (Rorty, 1979/2009). From the latter perspective, one could come to know matters of fact about the world simply by having clear perceptions of its given deliverances in sensory experience and, in turn, by representing these brute data in the form of 'protocol sentences'—the basic atomic constituents of meaning.[5] For the more foundationalist Vienna thinkers like Schlick, these observational reports, when testably accurate and formulated under the syntactic constraints of formal logic, were to provide empirical building blocks of the sort that uncovered basic extra-linguistic truths about reality. However, this downplayed the interpretive liabilities and potential errors of the human intermediaries and symbolic media through which that reality was inescapably filtered. More significantly, it started the epistemic aspiration of scientific reasoning and explanation down what Schlick called a "Cartesian road" of observational certainty (Schlick, 1934/1959, p. 220) that erroneously treated meaning and knowledge atomistically and in a social vacuum.

---

[5] There is controversy about the extent to which the logical positivists can be placed in the more traditional empiricist heritage that embraced this kind of atomistic epistemology, and indeed the school evolved over time and developed rifts between more conservative members like Schlick and more liberal or 'tolerant' ones like Neurath and Carnap, who had more holistic and coherentist leanings. Friedman (1999) presents an interesting alternative account that sets distance between the Vienna Circle and more traditional forms of empiricism. Colorful evidence of this central problem with the logical empiricist thinking and its reputation is given by Hollinger (2011): "Seated across from a skeptical John Dewey, exasperated over his own apparent inability to explain logical empiricism and to persuade Dewey of the doctrinal ecumenism of the *International Encyclopedia of Unified Science*, Otto Neurath stood up abruptly, raised his right hand as if taking an oath, and declared vehemently: 'I swear we do not believe in atomic propositions.' Whether Neurath's left hand grasped a copy of the Bible or of Isaac Newton's *Principia*, or only a wisp of startled air, we do know that Dewey was charmed by the gesture."

Per contra, as another group of deflationary thinkers that included Ludwig Wittgenstein (the mature), W. V. O. Quine, and Wilfrid Sellars as well as later neo-pragmatists like Hilary Putnam, Richard Rorty, and Donald Davidson would soon establish, in the reality of the social-interactive processes of research and hypothesis testing, protocol sentences and observational reports did not gain their meanings as semantic standalones, which facilitated interpretively unmediated contact points between thought and reality. Rather, they were always already inferentially interrelated to a manifold of other beliefs in the complex and ever-evolving meshwork of linguistic communication and cooperatively negotiated scientific understanding. By a similar token, the immediate and noninferential deliverances of brute data in the experimental circumstance could not themselves serve as *reasons* to support a particular theory (as had been implied in the logical empiricist's epistemological reliance on unthematized observational yields), for any and all empirical warrant occurred in an inferentially articulated sphere of justification, what Sellars, affirming Kant, called a "space of reasons." Even the basic determination and definability of empirical referents depended, in this holistic light, not on the privileged relationship of word and object but rather on the interactive happenings of a public milieu where the contentful reckonings of definition could occur only amid the common criteria established in contexts of shared social practices.

In the end, the members of the Vienna Circle who prioritized the building blocks of observational certainty and logico-mathematical architecture had erroneously taken scientific justification to be a univocal *way to truth*, a *way to 'the way the world is,'* as opposed to viewing it as a multivocal, fallible, and indeterminate *way of symbolic sharing* that was situated in the ongoing *communication, participation, and understanding of warm-blooded, interacting people.* Such an anchoring of scientific cognition on a bedrock of reductive and atomistic epistemic presuppositions secured the indefeasible objectivity of truth-claims only at the expense of a proper acknowledgement of the primitive roles played by social interaction in the establishment of sense and significance and by the situated, sociohistorically embedded, and corrigible character human perception, measurement, and understanding.

## 4. Situating Universalism, Pluralizing Methodology

While Merton's embrace of many of the critical and progressive ideas that had arisen in the Vienna Circle are explicit in his account of the ethos of science, the influence that the Circle's tendencies toward technmodernist scientism and epistemic reductivism come through much more subtly. In the first place, he refers to 'science' in a singular, generic, and undifferentiated way, that is, as a kind of blanket term presumably meant to run the whole gamut of sciences (i.e., the physical sciences, the life sciences, the social sciences, etc.) (Crane, 1972; Jacobs, 1987). However, notwithstanding the vagueness in this specification, he defines its institutional goal of "certified knowledge" and the technical norms that undergird its impersonal criteria of validity—its procedural prerequisites for "systematic," "sustained," "valid," "true prediction" of "objective sequences and correlations"—specifically in terms of the exact, physical sciences. This sort of technical objectivity, he writes, "precludes particularism," and secures "the impersonal character of science" that backs its universalism (Merton, 1942/1968, p. 606).

Yet, what remains indiscernible to Merton here is that scientism is itself a kind of particularism, though a disciplinary one. It prioritizes one method or procedure of scientific justification to the exclusion of others, and it does this in a way that therefore, in his phrase, "imposes particularistic criteria of validity." Instead of recognizing the importance and value of epistemological diversity across the physical, life, and social sciences, it reduces these multiple levels of epistemic analysis to one plane of rational acceptability and institutionalized knowledge production. Lost, in this respect, is a nonexclusionary and holistic impetus to methodological pluralism and interdisciplinarity (Norgaard, 1989; Wildemuth, 1993). This more inclusive approach places the wide range of disciplines *epistemically on par*, enabling an appreciation and integration of the advancements in scientific knowledge provided, for instance, by understandings of the complex functional and teleological relationships within and between living organisms struggling to make their way in changing *umwelten* or by reconstructions and interpretations of the subjective meaning complexes of intentional action that pattern the social behaviors of symbol-bartering human beings. A universalism that cannot (or is not willing to) accommodate a disciplinary plurality of knowledgeable voices that may contribute to richer scientific understandings of any given problem ceases to be a universalism per se.

Besides impeding the scope of the imperative of universalism for which he is arguing, the disciplinary strictures of scientism also come into conflict with Merton's imperative of "organized skepticism"—the "methodological and institutional mandate" that scientists actively suspend their judgments and continuously engage in a critical and detached scrutiny of beliefs and validity claims. Though this norm is an undeniable sponsor of the salutary fallibilism that characterizes the modern scientific method and inductive reasoning, more generally, it ranges beyond methodological confines, demanding unbounded scrutiny, reflexivity and self-questioning. Thus, as Merton writes, "the scientific investigator does not preserve the cleavage between…that which requires uncritical respect and that which can be objectively analyzed" (Merton, 1942/1968, p. 614). However, when the exact, natural sciences become the "*scientia mensura,*" (i.e., Sellars's idea that "in the dimension of describing and explaining the world, science is the measure of all things, of what is that it is, and of what is not that it is not") (Sellars, 1956, p. 303), it is precisely this sort of "uncritical respect" (Merton, 1942/1968, p. 614) that is demanded from epistemically subjacent disciplines and perspectives. Such a demand runs counter to the departure-point axiom of interdisciplinarity: *de omnibus dubitandum est,* doubt everything, take no systems, categories or criteria as given, question the primitives as such. As skeptical as this maxim may sound, such a methodical and illimitable "irritation of doubt"[6] has, in fact, been at the root of paradigm-shifting advancements across all natural and human scientific disciplines, and it motivates the uptake of the adjacent norms of the ethos of science like inclusive participation and egalitarian consensus-building.

## 5. Parting With the Scientific "View from Nowhere" by Strengthening Objectivity and Empiricism

A final set of issues emerges from Merton's equivocal equation of universalism and technical objectivity with the "impersonal character of science." While it is undeniably essential that objective criteria of validity and procedural mechanisms of justification should secure intersubjectivity rather than revert to personal or political authority, this does not necessarily imply that science itself has an "impersonal character." A view of the ethos of science that brands it as 'impersonal' is liable to slide into a scientific 'view from nowhere' that is caught unawares of its own positionality and contextual conditions of possibility—an operational ignorance to the situated nature of learning, culture, and knowledge that runs the risk of severely impoverishing any aspiration to meaningful objectivity. To be sure, scientific observers and cognizing subjects are always already embedded in the systems they are observing, investigating, and explaining (Stalder, 2019; Von Foerster, 2003). They themselves occupy the problem domain no matter how big or small it is, and their interests and concerns inexorably shape the objects, instruments, and spaces of inquiry through which they carry out their own knowledge practices. The data they collect are therefore not given as such (recall that, in Latin, *datum* simply means 'what is given') but are rather intimately intertwined with who, how, and why they were collected.

---

[6] "Doubt is an uneasy and dissatisfied state from which we struggle to free ourselves and pass into the state of belief; while the latter is calm and satisfactory state.... The irritation of doubt causes a struggle to attain a state of belief. I shall term this struggle *Inquiry*" (Peirce, 1877/1955, p. 10).

It follows that observations, statements of fact, inferential claims, and epistemic techniques are all, in their own significant ways, inseparable from the social and agential position whence they came (Alcoff, 1988; D'Ignazio & Klein, 2020; Hall, 1990; Henderson, 2011; Reiter, 2018; Stalder, 2019).

Still, it is important to note that this acknowledgement of the situated character of scientific knowledge need neither spiral into epistemic relativism nor jump the guardrails of methodological pluralism to stake a claim to epistemological ascendancy. As Donna Haraway has put it, it is feasible "to have simultaneously an account of radical historical contingency for all knowledge claims and knowing subjects, a critical practice for recognizing our own 'semiotic technologies' for making meanings, and a no-nonsense commitment to faithful accounts of a 'real' world" (Haraway, 1988). In other words, reflexive and context-aware scientific practice can retain a commitment to objectivity and universalism even if it molts the skin of its disembodied, "impersonal character." Indeed, thinkers like Sandra Harding have made the even more robust claim that adopting a "standpoint epistemology," which is cognizant of how everyday dynamics of bias and marginalization seep into all dimensions of research and innovation practices, can yield "strong objectivity" (Harding, 1992, 1995, 2008, 2015). This type of objectivity, she argues, is fortified through positional self-awareness and critical and systemic self-interrogation and hence ends up being more objective and more universalistic than weak modes of scientific or technical objectivity that stake an unobstructed claim to impersonality, neutrality, and value-free knowledge work. Rather than "valorizing the neutrality ideal" (D'Ignazio & Klein, 2020, p. 82) or peddling the "myth of neutrality," as Leonelli calls it, this sort of "strong objectivity" starts from a reflective recognition of the differential relations of power and social domination and then actively endeavors to include marginalized voices in the community of inquiry in order to transform situations of social disadvantage into situations that are scientifically and epistemologically richer and *more advantaged*. Such richer and more inclusive epistemic ecologies end up producing more comprehensive knowledge and more just and coherent practical and societal outcomes (Collins, 2017; Harding, 2008; Stoetzler & Yuval-Davis, 2002).

This first-order orientation of scientific investigation to societal betterment and social justice uncovers one last ambiguity in Merton's appeal to the impersonal character of science that merits exploration. Merton links the trait of impersonality closely with what he calls the imperative of "disinterestedness"—the institutional norm that guarantees the "public and testable character of science" through the integrity of scientists and their "ultimate accountability to their compeers" (Merton, 1942/1968, p. 614). Disinterestedness, he emphasizes, should not be confused with an "altruistic concern for the benefit of humanity," for this would run together institutional and motivational levels of analysis and erroneously insert the interests of the scientist where the constituent mechanism is, in reality, "a distinctive pattern of institutional control" (Merton, 1942/1968, p. 613).

Again, here, Merton seems to run the risk of veering into the ditch of a scientific 'view from nowhere.' An impersonal, agentless characterization of the norm of disinterestedness as a pattern of institutional control leaves one to suppose the existence of the institutional pattern, without further ado, while it neglects to provide an account of how and from where the pattern arises. In a similar fashion, it fails to explicate the role played by scientists as cognizing and concerned subjects who establish the changing space of normative parameters around their knowledge practices, who set the direction of travel for the activities of research and innovation, and who hold each other accountable by engaging in a reciprocal commitment to provisionally and indefinitely follow the force of good reasons. None of these aspects of science as social action and interactive praxis is "disinterested," properly speaking.

To respond to this deficit in Merton's functionalist account of the norm of disinterestedness, we may do well to take on what Bruno Latour has termed a "stubbornly realist attitude" (Latour, 2004, p. 231). Latour's realism calls into question the independent status of "matters of fact" inasmuch as these are conceived to be "given in experience" and then confronted by neutral and disinterested scientists who are charged to investigate them (Latour, 2004, p. 232; Stalder, 2019). What is needed instead, for Latour, is a stronger empiricism that views matters of fact as very incomplete and subsidiary renderings of what, in actuality, are *matters of concern*. Matters of fact might show up as natural phenomena, measurements, or data sets but these are, at the same time, taken up by scientists as matters that bear significance and that are deemed worthy of attention, care, investigation, and intervention. A stronger objectivity alive to the positionality of scientific subjects demands this kind of stronger empiricism that registers both the instantiation of sociality in its objects and the inscription of care in their study, use, and materiality (de la Bellacasa, 2017)—one that does not presuppose a universe of scientific investigation that is conceptually carved out in advance into an independent empirical realm of facts and data, an ideal realm of logic and language, and a neutral scientific arbiter trying to disinterestedly stitch these together (Davidson, 1973).

## 6. Unbounding Communalism

In spite of the shortfall of Merton's institutional imperative of disinterestedness in accounting for the social good orientation of scientific practice, another of his norms—what he calls the "communism of the scientific ethos"—goes a way to making more explicit the normative footing of science's greater "concern for the benefit of humanity." Thinking with Merton against Merton, this communitarian norm supplants the impersonal character of disinterestedness with a broader solidaristic commitment: "The substantive findings of science are a product of social collaboration and are assigned to the community" (Merton, 1942/1968 p. 610). This creates a "communal character of science" that is "further reflected in the recognition by scientists of their dependence upon a cultural heritage to which they lay no differential claims. Newton's remark—'If I have seen farther it is by standing on the shoulders of giants'—expresses at once a sense of indebtedness to the common heritage and a recognition of the essentially cooperative and selectively cumulative quality of scientific achievement" (Merton, 1942/1968, p. 612). But, beyond creating a solidarity between past and present participants in the broader human project of scientific inquiry, this intergenerational self-understanding also points to the future-oriented purpose of 'advancing the boundaries of knowledge.' The common possession of science, together with its goal of advancing knowledge that benefits all, institute it as an inviolable part of the 'public domain.' This public bearing requires of all scientific endeavors full, open, and unbounded communication that enacts "the moral compulsive for sharing the wealth of science" and makes those who do not share the public property of their scientific genius guilty of being "selfish and anti-social" (Merton, 1942/1968, p. 611).

Although Merton's inclusion of the norm of communalism as an "integral element of the scientific ethos" adumbrates that there is a kind of moral grammar that underlies the public-good orientation of concrete scientific practices, his reconstruction focuses primarily on the rationale behind the common ownership of science rather than on the normative structures that subtend its institutional commitment to societal betterment. It is the American pragmatist thinker, Charles Sander Peirce, who labors to provide a viable blueprint of these moral-practical structures. For Peirce, an orientation to the betterment of humanity is an intrinsic feature of the logic of scientific inquiry. The scientific community of inquirers engages in an intergenerational and indefinite search for the truth, which aims at an "ideal limit" point of general consensus "towards which endless investigation would tend to bring scientific belief" (Peirce, 1901/1931, 5.565). In other words, true beliefs, for Peirce, are those that would withstand scientific scrutiny were inquiry to go on infinitely (Capps, 2019). For this reason, as Justus Buchler puts it, "to Peirce the scientific method represents the antithesis of individualism" (Peirce & Buchler, 1955, p. x). Mortal scientists dedicate their lives to a search for truth that is, by its very nature, out of their reach and to a mode of scientific progress, societal betterment, and species advancement the fruits of which only future generations of citizens, subjects, and scientists are likely to fully reap. In an essay titled, "Grounds of the Validity of the Laws of Logic," Peirce points out that the very fact that scientists' concerns for the future far outdistance their lifetimes reveals the rationality of "complete self-sacrifice" that pilots their dedication and "the logical necessity of complete self-identification of one's own interests with those of the community" (Peirce, 1869/1931, 5.356; Schlaretzki, 1960; Ward, 2018). It is, in fact, precisely the unscalable chasm between the finite scientific investigator and the infinite, temporally unbounded community of inquirers—as well as the unbounded 'we' of the future human community as a whole—that leads Peirce to see the logic of scientific inquiry, first and foremost, "as a species of ethics" (Legg, 2014). "Logicality," he writes, "inexorably requires that our interests shall not be limited. They must not stop at our own fate, but must embrace the whole community" (Peirce, 1910/1955, p. 162). "Logic is rooted in a 'social principle,' for investigation into what is true is not a private interest but an interest 'as wide as the community can turn out to be'" (Peirce, 1910/1955, p. 162; Peirce, 1931, 5.357; see also Misak, 2012).

Peirce's allusion to a common interest "as wide as the community can turn out to be" should have special significance for a contemporary community of inquiry doing science in the age of the Anthropocene. The multiplying species extinction risks and threats to biospheric destruction that derive from applied scientific activity are ever more compelling scientists to see themselves as part of a broader biotic totality, part of the unbounded community of the biospheric sum (Leslie, 2020, Appendix). And, it is precisely this compulsion that nudges researchers to think beyond themselves and beyond their existing communities of practice to consider their roles in safeguarding the endurance of a greater living whole that gives special relevance to Peirce's mining of the moral grammar underlying the public-good orientation of concrete scientific practices. Peirce anticipates this wider dimension too. He writes of the logicality behind the "infinite hope" that pursues the "supreme and transcendent interest" of sustaining life in the face of threats to its total annihilation (Peirce, 1869/1931, 5.357). And, he argues that members of humanity must move beyond the psychologistic and anthropocentric barriers they set up between themselves and the natural world so to embrace their continuity with life and the agapic architectonics of self-sacrifice that have underwritten the endurance of species since time immemorial (Anderson, 1995; Burch, 2018). To this end, he urges that scientists and civilians alike assume a kind of rational responsibility to the expanding circle of life: "This community, again, must not be limited, but must extend to all races of beings with whom we can come into immediate or mediate intellectual relation. It must reach, however vaguely, beyond this geological epoch, beyond all bounds. He who would not sacrifice his own soul to save the whole world, is, as it seems to me, illogical in all his inferences, collectively" (Peirce, 1910/1955, p. 162).

His hyperbole notwithstanding, Peirce's extension of the normative inner logic of science's commitment to societal wellbeing and betterment to a spatially and temporally unlimited community anticipates the crisis now faced by the contemporary natural and human sciences. In present global society, the technological enhancement of humanity's practical efficacy (i.e., its potential to enact global catastrophe, extinction, and ecological harm) has dangerously outpaced corresponding learning processes in the development of social, ethical, and political self-understandings capable of responsibly managing and constraining such hazardous potentials. The species-level, intergenerational, and biospheric challenges now posed by the impacts and the exponential acceleration of scientific advancement have created an urgent need, across all domains of learning and culture, to widen the moral-practical purview of humankind's approach to research, innovation, and discovery. They have created a need to reconceptualize and rescale the primitive categories of social, ethical, and political analysis—to think beyond the nearsighted values of growth, efficiency, and utility-maximization, beyond consequentialist calculi that hem the pursuit of interests into a myopic here-and-now, beyond the vicissitudes of statist and nationalist hubris that have created multiplying incentives to engage in an internecine geopolitical race to a nonexistent technological finish line. Such a felt demand to confront the universal challenges of the society of tomorrow by overhauling the conventional vocabularies of today requires the collaborative creation of novel justificatory conduits by means of which biospheric and species-level solutions can be suitably formulated, considered, and assessed. More than this, they require a commitment to the global, indeed biospheric, public good, of the sort that Peirce finds at the very heart of science itself.

## 7. Understanding the Ethos of Responsible and Sustainable Data Work

I have gone to great lengths to put Merton's idea of the normative structure of science into dialogue with both its intellectual influences and critics so that I could conclude by capturing the implicit normative advancements Leonelli makes in her article. Indeed, critically foraging through Merton's norms, we start to better discern the scaffolding of normative assumptions that supports Leonelli's implicit appeal to the values of epistemic integrity and the global public good that work together to animate the ethos of responsible and sustainable data work. Her appeal to epistemic integrity—what I have described as the commitment to provide as full and inclusive yet reflexive and critical a data scientific understanding as possible—lines up with the demand to *situate universalism* in a methodologically pluralistic, interdisciplinary, and critical approach to subject-stationed and socially embedded scientific understanding. It also aligns with the charge to *strengthen objectivity and empiricism* through a responsiveness to context, a reflexive awareness of positionality, and an unflinching inclusivity that labors to register the expertise and lived experience of all affected individuals and communities. Her appeal to the global public good and to the stewardship of planetary health lines up with the imperative to overcome the lures of impersonality and neutrality through transformative knowledge work dedicated to societal betterment and social justice, and it tracks the impetus to *unbound communalism* so to extend the commitment to carry out the endeavors of science in the public interest to the unbounded community of the biospheric whole.

To take each of these norms in order, Leonelli's rendering of *situated universalism* explicitly eschews the "misguided and autocratic view of science" that excludes "social, contextual factors from evidential reasoning, and thus [disregards] the conditions under which data are generated and interpreted." This latter type of ersatz universalism produces data work that is not fit-for-purpose in responding to the complex societal and public health problems that arise in times of pandemic. An epistemological gambit that attempts to tackle such problems through quantitatively and computationally anchored explanation or prediction alone will simply be inadequate to account for the broad-ranging causal and qualitative factors—such as the social determinants of infectious disease vulnerability or the "perpetuating configurations of noxious social conditions" (Bambra et al, 2020, 965) behind the pandemic's disparate effects—that should inform responsible, sustainable, and effective public health decision making and clinical action. The *explanans* here (namely, logically sound code and featurized data) does not provide sufficient grounds for the *explanandum*. That is, the methods, evidence, and inference types used to analyze the phenomenon are not adequate for grasping the phenomenon under analysis. This is why the epistemic integrity of scientific understanding is so crucial. It introduces a "third term" to the impersonal relation of the *explanans* and *explanandum* (de Regt et al., 2009), namely, cognizing subjects, who can use their contextually aware judgment to include a range of explanatory and interpretive paths in holistically tackling a problem. This more situated but more universalistic approach to scientific understanding enables a diverse range of *explanantia* to be woven together and made properly responsive to the complexities and contextual nuances of their *expalananda*. A situated universalism accomplishes the latter by taking on an 'ecosystem view' of data science (Meng, 2019), which, as Leonelli argues, "sees this domain as a catalyst for contributions from several different disciplines, including both STEM [science, technology, engineering, and mathematics] and SHAPE [social sciences, humanities and the arts for people and the economy] subjects."

It is worth emphasizing here that the perspectival inclusiveness of *situated universalism* should be viewed as a distinctive and vital element of any mode of applied data science or artificial intelligence innovation that is discharged to address societal challenges. Exceptionally, these approaches are entrusted to serve surrogate cognitive functions in the human world. They 'stand-in,' as it were, for different tasks that would otherwise require the interventions of thinking, judging, and inferring human subjects, who use their perceptual knowledge, logic, understanding, common sense, social wherewithal, and moral compass to navigate the unbounded problem domains of shared experience (Leslie, 2019, pp. 39–40). This fiduciary, stand-in role comes at a high epistemic price: Insofar as data scientific and AI applications do things that require intelligence when done by humans—insofar as they function to address problems that would otherwise require *judgment, reasoning and interpretation*—they also require a dispatch of the same array of cognitive resources as those upon which humans draw in their own practical dealings with social reality. These resources are, at least, four-fold. They include aspects of *logic* (applying the basic principles of validity that lie behind and give form to sound thinking and acting), aspects of *semantics* (gaining an understanding of how and why things work the way they do in the real world and what they mean), aspects of the *social understanding of practices, beliefs, and intentions* (interpretively clarifying the content of interpersonal relations, cultural and political structures, societal norms, and individual objectives), and aspects of *moral reasoning and justification* (making sense of what should be considered right and wrong in our everyday activities and choices). Responsible and sustainable data work that addresses societal problems must include and integrate all of these cognitive trajectories. It should be logically sound, interpretable, coherent, and evidence-based but also sufficiently responsive to social, cultural, and political contexts and ethically robust. The commitment to epistemic integrity requires such a perspectivally inclusive, interdisciplinary, methodologically pluralistic, and integrative 'ecosystem view.'

As has become clear, the norm of *situated universalism* significantly joins up with the imperative of *strengthening objectivity and empiricism* through contextual responsiveness, reflexive awareness of positionality, and an unwavering inclusivity. Critically, however, the normative demands of *strong objectivity and empiricism* move beyond the more epistemic focus of *situated universalism* in an important way. The inclusivity dimension of the former refers not just to the perspectival inclusivity needed for methodological pluralism but also to a participatory inclusivity that ensures that data work incorporates horizontal public engagement, community-involved co-production, and social mobilization. In keeping with this more substantive and democratically informed valence of inclusivity, researchers are obliged to start from a standpoint of positional self-awareness that situates their own research activities and communities of practice in the contexts of social stratification and differential relations of power in which they are embedded. This then allows them to reflectively safeguard that the voices, lived experience, insights, and needs of impacted individuals and communities—especially those of marginalized groups—are properly heard and heeded. In this connection Leonelli calls "for researchers to move away from data collection as a top-down exercise in surveillance, and toward collaborative, engaged forms of data work that seek to understand social and environmental needs, evaluate research directions and construct appropriate tools in dialogue with relevant communities. Community engagement is crucial to obtaining robust data as well as robust data use and outputs."

Following from this investment in safeguarding the material conditions necessary for the realization of equitable participation in knowledge production, responsible and sustainable data work carried out with a commitment to *strong objectivity and empiricism* takes up problem-solving tasks in an unapologetically nonneutral and publicly oriented way. It sees its objects not as inert heaps of observational or experimental data awaiting the technical virtuosity of shrewd analysts armed with toolboxes of algorithmic techniques, but rather as matters of concern to be confronted through engaged, holistic, and transformative knowledge practices that are dedicated to social justice and societal betterment.

If anything, the COVID-19 pandemic has shown that the excrescences of technical virtuosity and technocratic aspiration, which predate the outbreak but continue on unabated, have led, in many cases, to a downward spiral. Exacerbated as it has been by the algorithmically supported dispersal of misinformation at scale (Burki, 2020; Centre for Countering Digital Hate, 2020) and debilitating levels of digitally supercharged social and political polarization (Cinelli et al., 2020; Parmet & Paul, 2020), the pandemic has also ushered in myriad forays into questionable and scientifically tenuous rapid-response predictive analytics (Wynants et al., 2020) and ineffective health surveillance technologies that have circumvented society-centered and trust-generating public health practices while opening up real possibilities for long-term function creep (Leslie, 2020). Likewise, it has summoned the hasty repurposing of data scientific applications and AI systems that are uniquely positioned to further entrench, exacerbate, and augment existing societal inequities and health inequalities (Leslie et al., in press). This is well illustrated in the U.S. prison system's use of a discriminatory recidivism prediction tool with a documented history of racial bias to determine which inmates are released to home confinement, thereby exposing African American inmates to increased COVID-19 exposure and heightened health risk (Fogliato et al., 2020). Questions abound in all these cases as to whether a more self-aware starting point in public-good oriented, engaged, and transformative data work that abandons

"the myth of neutrality" would have led the researchers and technologists behind these interventions down different, more responsible paths. What is undeniable, however, is that the ethos of responsible and sustainable data work demands that data practitioners reflexively engage with the commitments and values that are driving their efforts, so that they can assess the degree to which these are yielding socially beneficial solutions and assume ethical ownership for their near-term consequences and long-term impacts. As Leonelli argues, "it is imperative that data scientists take responsibility for their role in knowledge production."

Already implied in the norm of nonneutral, publicly oriented data work is a commitment to serve the interests of the community, but this is not yet what I have called *unbounded communalism*. For Leonelli, responsible data science in times of pandemic and beyond must rescale the spatial and temporal parameters of its public interest orientation to accommodate the planetary scope of public health crises, the biospheric reach of anthropogenic climate change, and the multiple interdependencies and possibilities that have been created by global digital transformation. This rescaling involves both a *looking outward* and a *looking forward*. That is, rather than looking inward and focusing on how to leverage quick-fixes and cull insights from national-level data, responsible and sustainable data work must look outward to building transnational research cooperations and data infrastructures that are equitably resourced and capacitated (Bezuidenhout et al., 2017) and that foster "a localized, situated, procedural understanding of the conditions and behaviors most likely to stem [disease] transmission and improve (not just human, but planetary) health." Likewise, it must look forward to developing 'longer-term analyses' that work to understand the broader societal and planetary implications of potentially short-sighted technological interventions. By building "research incentives toward longer term, complex solutions," into pandemic response strategies, these can be intertwined with "ongoing efforts to address the existential threats posed by climate change" and to steward long-run planetary wellbeing. Bearing in mind this global and biospheric stretch of Leonelli's vision of the ethos of responsible and sustainable data work, it may well be said that its communalism is, in no uncertain terms, unbounded.

## 8. Conclusion

The arc of the data scientific universe is long, but it bends toward the global public good. Though the obtainment of this end is as yet not widely realizable in practice, though we are still at the beginning point of a steep incline on the hither side of a hill whose upward curvature seems incalculable, the ethos of responsible and sustainable data work comprises the sturdy frame of the moral arc upon which data scientists are already starting off on that climb. This image of a moral arc is derived, of course, from the saying of Martin Luther King, Jr.: "The arc of the moral universe is long, but it bends toward justice." Imagining what he called a "forward stride toward the city of freedom," beyond "rocky places of frustration…and the fatigue of despair," King urged that the ultimate rise of justice in the world give those fighting racism and oppression "courage to face the uncertainties of the future" and the strength to overcome domination (King, 1967/2001). King's imagery of the arc drew upon the writings of the great abolitionist, Theodore Parker, who in 1853 had written:

> Look at the facts of the world. You see a continual and progressive triumph of the right. I do not pretend to understand the moral universe, the arc is a long one, my eye reaches but little ways. I cannot calculate the curve and complete the figure by the experience of sight; I can divine it by conscience. But from what I see I am sure it bends towards justice." (Parker, 1853)

King's and Parker's imaginaries are a form of what the philosopher Joshua Cohen has called "ethical explanation," namely, the putting forward of ethical norms—such serving the ends of justice or the global public good—as an elucidation of why some social fact or outcome can be expected to obtain (Cohen, 1997, 2010). For Cohen, ethical explanations are not necessarily "simply collages of empirical rumination and reified hope, pasted together with rhetorical flourish," but rather plausible forms of argument where the iniquity or injustice of a given social arrangement limits its viability for reason of the voluntary and cooperative character of the long-term sustainability of that arrangement (Cohen, 1997, p. 93). On this view, "social arrangements better able to elicit voluntary cooperation have both moral and practical advantages over their more coercive counterparts" (Cohen, 1997, p. 93).

This especially true for responsible and sustainable data work, for its moral and practical advantages are equiprimordial and convergent. It is not just that *good data science* is *both* robust, reliable, and sound data science *and* publicly oriented data science—*data science for the social good*. Beyond this, there is the stronger claim that good data science is good *because it is for the social good*. Its moral qualities are, in actuality, conditions of possibility for its practical success, or to put it slightly differently, the moral qualities that allow for it to serve the social good—those constituents of its ethos like inclusive and equitable participation, unbounded, universalistic solidarity, pluralistic and situated universalism, dialogical openness, institutionalized criticism, and strong, contextually responsive objectivity oriented to social justice—are requisite ingredients of its epistemic and practical viability. Correspondingly, iniquities and injustices in the sociotechnical arrangements behind its conditions of knowledge production (such as entrenched biases, discriminatory exclusions, scientistic particularism, and technocratic myopia) severely limit its long-term sustainability.

Be that as it may, we should note that to say that the arc of the data scientific universe bends toward the global public good is not to say that the journey to the realization of this end is inevitable or predetermined. Ethical explanation should not generate the complacency or quietism that may arise as a compliment to the sense that a better future lies ineluctably in wait. Data science can cease to be scientific, in the proper and widest sense of that word, when fleet-footed ambition, parochialism, self-interest, corporate power, and legacies of socio-political domination lead it astray from its native social, practical, community-oriented, and universal path. Indeed, an awareness of its moral arc enjoins us, *as scientists*, to confront such aberrations head-on and to face up to the real-world barriers that stand in the way of its actualization. It enjoins us to interrogate the material preconditions that are necessary for its realization and to bring these about.

This becomes especially critical in a contemporary sociotechnical reality where, despite the current swell of appeals to data science and AI 'for the social good' in the academic literature (Berendt, 2019; Floridi et al., 2020; Hager et al., 2019; Shi et al., 2020; Taddeo & Floridi, 2018), there have emerged multiplying extractive constellations of "data colonialism" (Magalhães & Couldry, 2020) and "philanthro-capitalism" (Burns, 2019). Notwithstanding their projections of charitable rhetoric, such patterns of the "commercially-driven production of the social good" (Magalhães & Couldry, 2020) are heralding the convergence of high-entry-cost digital innovation ecosystems and multinational corporate business models and marketing strategies with humanitarian data work, development schemes, and market forces, thereby "materializing discrimination associated with colonial legacies…[and] contributing to the production of social orders that entrench the 'coloniality of power'" (Madianou, 2019; Mohamed et al., 2020; Quijano, 2000). These trends are, in turn, signaling "a profound rebalancing of power and governance in the domain of social life, privileging corporations with large-scale data power and making states (and other commercial and civil society actors) dependent on those corporations" (Magalhães & Couldry, 2020). Similarly, incipient forms of data governmentality and commodification are ever more infiltrating academic venues and research environments where large tech companies—armed with proprietary data sets, seemingly unlimited financial resources, and massive computing power—are, in effect, 'de-democratizing' AI and data science (Ahmed & Wahed, 2020; Van Dijck, 2018). This is occurring, inter alia, through their control over access to data and compute resources, their command over labor power by means of the university-corporate hybridization of 'dual-affiliation' career trajectories (Roberge et al., 2019), and their manipulation of the terms of open research to protect their own rentiership claims to monopolistic control over intellectual property and infrastructural assets (Abdalla & Abdalla, 2020; Amodei & Hernandez, 2018; Birch, 2020; Birch et al, 2020; Frank et al., 2019; Gupta et al., 2015; Lohr, 2019; Riedl, 2020; Roberge et al., 2019).

From this more cautionary (yet more objective and empirical) vista, we are compelled to ask difficult questions: Is the current universe of data scientific innovation—that is, the global assemblage of data infrastructures, compute infrastructures, algorithmic infrastructures, funding schemes, and research and delivery capabilities and resources—equipped to actualize the sort of globally minded, responsible, and sustainable data work toward which the moral arc of data science bends? Or, do these infrastructural, resourcing, and human factors operate, in fact, to curtail and elide possibilities for that actualization? Is the human and biospheric interest in the realization of the global public good being served by the current constellation of power relations, sociotechnical affordances, platform ecosystems, and infrastructural networks that characterize contemporary data scientific research and innovation environments both nationally and globally? Such questions suggest that rather than training our sights on the long arc of the data scientific universe, trying to portend its ultimate designs on a better, more ethical future, we must first and foremost look critically at and around ourselves, in the vexing mirror of the present, so that we can scrutinize the systemic inequities and structural injustices that delimit the knowledge practices of today. As Alan Turing would say, "we can only see a short distance ahead, but we can see plenty there that needs to be done."

Cohen, J. (1997). The arc of the moral universe. *Philosophy & Public Affairs*, *26*(2), 91–134.

Cohen, J. (2010). *The arc of the moral universe and other essays*. Harvard University Press.

Collins, P. H. (2017). Intersectionality and epistemic injustice. *The Routledge Handbook of Epistemic Injustice,* 119. Routledge. https://doi.org/10.4324/9781315212043-11

Crane, D. (1972). *Invisible colleges*. The University of Chicago Press.

Davidson, D. (1973). On the very idea of a conceptual scheme. *Proceedings and Addresses of the American Philosophical Association, 47*, 5–20. https://doi.org/10.2307/3129898

de La Bellacasa, M. P. (2017). *Matters of care: Speculative ethics in more than human worlds* (Vol. 41). University of Minnesota Press.

De Regt, H. W., Leonelli, S., & Eigner, K. (Eds.). (2009). *Scientific understanding: Philosophical perspectives*. University of Pittsburgh Press.

D'Ignazio, C., & Klein, L. F. (2020). *Data feminism*. MIT Press.

Elster, J. (2009). *Alexis de Tocqueville, the first social scientist*. Cambridge University Press.

Floridi, L., Cowls, J., King, T. C., & Taddeo, M. (2020). How to design AI for social good: Seven essential factors. *Science and Engineering Ethics*, 26(3), 1771–1796.

Fogliato, R., Xiang, A., & Chouldechova, A. (2020). *Why PATTERN should not be used: the perils of using algorithmic risk assessment tools during COVID-19*. Partnership on AI. 2020.
35